\documentclass[showpacs,preprintnumbers,amsmath,amssymb]{revtex4}
\usepackage{graphicx}% Include figure files
\usepackage{dcolumn}% Align table columns on decimal point
\usepackage{bm}% bold math
\usepackage[T1]{fontenc}
\usepackage{ae,aecompl}
\begin{document}

%\title{Second order {\it cell} strong coupling perturbative expansions for superlattice Bose-Hubbard models}
\title{{\it Cell} strong coupling perturbative approach to the phase diagram of ultracold bosons in optical superlattices}

\author{P. Buonsante}
\affiliation{Dipartimento di Fisica, Politecnico di Torino and I.N.F.M, Corso Duca degli Abruzzi 24, I-10129 Torino (ITALIA)}%
\author{A. Vezzani}
\affiliation{I.N.F.M. and Dipartimento di Fisica, Universit\`a degli Studi di Parma, I-43100 Parma (ITALIA)}
%\email{cassi@fis.unipr.it}

\date{\today}% It is always \today, today,

\begin{abstract}
The phase diagram of ultracold bosons in realistic optical superlattices is addressed via second-order {\it cell} strong coupling perturbative expansions for the Bose-Hubbard model describing the system.
%We address the superfluid-insulator quantum phase transition of the Bose-Hubbard model describing ultracold atoms in one-dimensional realistic optical superlattices via a second-order {\it cell} strong coupling perturbative expansion (SCPE). 
Taking advantage of the cell partition inherent in the complex periodic modulation of a superlattice, this technique allows for the description of  the unusual loophole-shaped insulator domains that may occur in the phase diagram of the system, unlike the standard perturbative approach.
Furthermore, comparisons with quantum Monte Carlo simulations show that our approach  provides quantitatively satisfactory results at a significantly lower computational cost than brute force numerical methods. 
We explicitly consider the phase diagrams for two realistic 3-periodic optical superlattices. These show that many insulator domains exhibit an unusual reentrant character, which we discuss, and suggest that the quantum phase transition relevant to the loophole-shaped insulating domains does not require extreme experimental condition in order to be observed.
\end{abstract}


\maketitle

\section{Overview}
 In a recent experiment \cite{A:Peil}, Bose-Einstein condensates (BECs) were loaded in a one-dimensional trapping potential created by  superimposing two independent one-dimensional homogeneous optical lattices with commensurate lattice constants $d_1 = \ell_1 a$ and $d_2 = \ell_2 a$, $\ell_j \in {\mathbb N}$ \cite{N:OL}.  This gives rise to an optical trapping potential characterized by a periodic modulation involving $\ell = \max(\ell_1,\ell_2)$ subsequent local minima, i.e. a $\ell$-periodic superlattice. 

 In this paper we address the superfluid-insulator quantum phase transition that is expected to characterize this system, due to the competition between the boson repulsive interaction and the kinetic energy \cite{A:Fisher}. We recall that the latter can be tuned experimentally, being directly related to the strength of the optic potential \cite{A:Jaksch}. The control of this parameter actually played a key role in  recent experiments where the superfluid-insulator quantum phase transition was revealed for two and three dimensional arrays of Bose-Einstein condensates \cite{A:Greiner02,A:Kohl}.

In particular, we provide the zero-temperature phase diagram for the Bose-Hubbard (BH) model, originally introduced for liquid helium in porous media \cite{A:Fisher} and nowadays commonly adopted in the description of BEC arrays trapped in optic lattices \cite{A:Jaksch} and superlattices \cite{A:Roth03,A:Santos,A:loophole}. We recall that in the latter case the phase diagram may exhibit  unusual loophole shaped insulator domain corresponding to fractional fillings. These domains, that  cannot be described by standard  strong coupling perturbative expansions (SCPE), have been shown to exist by exploiting the exact mapping between  hard core bosons and free spinless fermions on one dimensional systems \cite{A:loophole}, and subsequently investigated  by means of a qualitative multiple-site mean field approach \cite{A:LoopLP}. In the case of $\ell=2$ superlattices, quantitative results have been provided in terms of (analytical) first-order cell SCPE and quantum Monte Carlo simulations \cite{A:loophole}.

Here, such phase diagram are obtained in terms of second order {\it cell} SCPE for the ground state energy of the system, that --- unlike the standard SCPE --- allow a rather satisfactory quantitative description of all of the critical fractional fillings characterizing superlattices \cite{A:loophole}. 
% This is possible basically because in cell SCPE the ``building blocks'' of the unperturbed system are not individual lattice sites, as in standard SCPE \cite{A:Freericks2}, but rather  the individual unit cells inherent in the complex periodic nature of the superlattice.
This is possible basically because in cell SCPE the unperturbed system is not a collection of individual lattice sites \cite{A:Freericks2}, but rather the collection of the individual unit cells inherent in the richer periodic nature of superlattices. For all of the fractional fillings determining the key features of the zero-temperature phase diagram, this choice allows the analytic description of the unperturbed ground state that lies at the basis of a perturbative expansion.
%Hence,  for all of the fractional fillings determining the key features of the zero-temperature phase diagram, it is possible to provide the analytic description of the ground state of the unperturbed system that lies at the basis of a perturbative expansion.

The price for the improved description granted by cell SCPE is the loss of the completely analytical treatment allowed by standard SCPE. Indeed, completely  analytical results based on cell SCPE  are approachable only at first perturbative order, for small cell sizes and fillings \cite{A:loophole}. On the other hand, cell SCPE provides quite satisfactory results with a relatively small numeric effort, consisting basically in the exact diagonalization of the BH Hamiltonian relevant to the isolated unit cell of the superlattice. This allows a reliable quantitative description of situations beyond the reach of  standard SCPE and/or  quite demanding for more refined numeric approaches, such as quantum Monte Carlo simulations (relatively large fractional fillings for superlattices whose unit cell comprises several sites).

%The most relevant result evidenced by our cell SCPE concerns the  critical  value of the ratio of the kinetic to the repulsion energy for the loophole insulating domains. 
The most noteworthy result evidenced by our cell SCPE concerns the  critical kinetic to repulsive energy ratio characterizing the loophole insulating domains. 
Indeed,  according to previous  estimates based on a mean-field approach \cite{A:loophole}, the loophole insulating domains are entered at critical values of the above ratio much smaller than that relevant to the usual lobe-shaped domains. Our second-order cell SCPE, which provide a more quantitative estimate than mean-field approximations, show that the critical values for the loophole domains are of the same order, and sometimes even larger, than those of the usual lobe domains with comparable critical fillings. This suggests that the insulating phase corresponding to  fractional filling loophole domains does not require extreme experimental conditions. Furthermore, second-order cell SCPE confirm the reentrant character of the insulating domains predicted for superlattice by a mean-field approach and first-order cell SCPE \cite{A:loophole}.

The plan of this paper is the following. In Sec. \ref{S:BH} we recall the {\it superlattice Bose-Hubbard model} \cite{A:Roth03,A:Santos,A:loophole} for the description of ultracold atoms in optic superlattices and introduce the key ideas of cell SCPE, the details of this approach being confined in the appendix for the sake of readability. 
In  Sec. \ref{S:results} we illustrate  the capabilities of the technique by comparing  cell SCPE and quantum Monte Carlo results for a $\ell=2$ superlattice. We confirm that cell SCPE is satisfactory also when the perturbative parameter is not very small. In particular quite satisfactory results are obtained even in the the homogeneous case. Furthermore we provide  the zero-temperature phase diagrams for some realistic one dimensional superlattices \cite{A:Peil}. 
Section \ref{S:concl} contains our conclusions.
\section{Superlattice BH model and {\it cell} SCPE}
\label{S:BH}
 Following the tight-binding-like approach of Ref.~\cite{A:Jaksch}, ultracold atoms in optical superlattices are commonly described by means of a {\it superlattice Bose-Hubbard model} \cite{A:Roth03,A:Santos,A:loophole} characterized by a periodic modulation of Hamiltonian parameters such as the local energy offsets and the hopping amplitudes between neighbouring lattice sites.  
In the particular case of one dimensional $\ell$-periodic superlattices, a unit cell consists of $\ell$ subsequent sites, so that a site can be conveniently labeled by two indices, denoting the cell it belongs to and its position within such cell, respectively.
Hence, the Bose-Hubbard (BH) Hamiltonian for a superlattice comprising $C$ cells each containing $\ell$ sites can be written in the form
\begin{equation}
\label{E:BHH}
H = \sum_{c=1}^{C} H_c + V 
\end{equation}
where
\begin{equation}
\label{E:cBHH}
 H_c = \sum_{k=1}^{\ell} \left[ \frac{U}{2} n_{k,c} (n_{k,c}-1) -(\mu-v_k) n_{k,c}\right]  - \sum_{k=1}^{\ell-1}  t_k (a_{k,c} a_{k+1,c}^+ + a_{k,c}^+ a_{k+1,c})
\end{equation}
refer to the $c$th cell, whereas
\begin{equation}
\label{E:perT}
V = t_\ell \sum_c (a_{\ell,c} a_{1,c+1}^+ + a_{\ell,c}^+ a_{1,c+1})
\end{equation}
joins subsequent cells. Operators $a_{k,c}^+$, $a_{k,c}$ and $n_{k,c} =a_{k,c}^+ a_{k,c}$ create, annihilate and count bosons at site labeled $(k,c)$. As to the Hamiltonian parameters, $U>0$ is the boson (repulsive) interaction, $\mu$ is the chemical potential, $v_k$ is the energy offset of the $k$th site within the cell, the $t_k$'s  with $k = 1,2,\ldots,\ell-1$ are the intra-cell hopping amplitudes and $t_\ell$ accounts for the hopping of bosons across neighbouring cells. Of course on an infinite $\ell$-periodic  structure the $\ell$ possible choices for the boundaries of the cells are in principle equivalent. However, in view of our strong coupling perturbative expansion, the most reasonable choice is such that the inter-cell hopping amplitudes are the smallest.

As we  recalled above, the one-dimensional $\ell$-periodic superlattice Bose-Hubbard (BH) exhibits an insulator-superfluid quantum phase transition at any rational filling of the form $f = L/\ell$, $L\in {\mathbb N}$  \cite{A:Roth03}. These phase transitions are commonly put into evidence in the zero-temperature $\mu/U$-$t/U$ phase diagram, where $t$ is a scaling factor for the superlattice hopping amplitude related to the strength of the optic potential (that is $t_k = t \cdot \tau_k$). More to the point, the boundaries of the insulating domain relevant to the critical filling $f$ are obtained comparing the ground state energies of the systems containing $N = f M$, $N+1$ and $N-1$ bosons, where $M$ is the number  of sites in the lattice, which should be large enough to provide a satisfactory approximation of the thermodynamic limit. Since the size of the Hilbert space for $N$ bosons on $M$ sites is $d(N,M) = \binom{N+M-1}{N}$, exact  diagonalization is beyond the reach of common computational capabilities already for relatively small systems at low fillings, and a numeric estimate of the ground state energies under concern requires  complex algorithms such as quantum Monte Carlo \cite{A:Batrouni,A:Kashurnikov96} or density matrix renormalization group simulations \cite{A:Kuehner,A:Schollwoeck}.

In Ref.~\cite{A:Freericks2}, satisfactory approximations of the ground state energies required for the description of the insulator domains are obtained for homogeneous lattices in terms of analytic third order expansions in the ratio of the hopping amplitude and the boson interaction. This approach can be  extended to structures featuring inhomogeneous hopping amplitudes, including superlattices \cite{A:scpe}, but in general  for integer critical fillings only. Indeed, in the case of fractional critical fillings, an explicit form for the structure of the unperturbed ground state is might not be available for the boson populations involved in the determination of the insulator domain boundaries. This typically happens when some of the local potentials $v_k$ within the unit cell are equal, and loophole insulator domains appear in the phase diagram \cite{A:loophole}.
Consider for instance the extreme case where there is no modulation in the local potentials (i.e., without loss of generality, $v_k=0,\;\forall k$) and the periodic pattern of the superlattice is due to the hopping terms alone, which do not  appear in the unperturbed Hamiltonian of standard SCPE. The only fractional fillings for which the form of the unperturbed ground state can be easily provided are those differing by a single boson from an integer filling, i.e. those involved in the determination of the integer-filling insulator domains \cite{A:Freericks2,A:scpe}. This is basically due to the fact that the degeneracy of the lower  eigenspace of the unperturbed Hamiltonian equals the lattice size, so that finding the unperturbed ground state reduces to a trivial single-particle problem.
For any other fractional filling, the degeneracy of the lower unperturbed eigenspace is much larger than the lattice size, and the ground state cannot be found analytically. 

This problem can be circumvented adopting a perturbative approach that takes advantage of the periodic modulation of the Hamiltonian parameters characterizing a superlattice BH model, i.e. whose perturbative parameter is the inter-cell hopping amplitude \cite{A:loophole}. That is to say, in the {\it cell} SCPE the unperturbed Hamiltonian is the sum of the independent cell Hamiltonians
\begin{equation}
\label{E:H0}
H_0 = \sum_c H_c 
\end{equation}
while the perturbative term is that defined by  Eq.~(\ref{E:perT}).
We mention that a similar approach was adopted in the study of the Hubbard \cite{A:Bernstein74} and Heisenberg \cite{A:Yamamoto} models in the special case of a dimerized chain (i.e. a $\ell=2$ superlattice).
 In some sense, cell SCPE is  standard SCPE applied to the homogeneous lattice made up by unit cells instead of individual lattice sites. This choice makes it easy the determination of the analytic form of the unperturbed ground state for the total populations $N = C L$, $N= C L\pm 1$ relevant to the  fractional critical filling $f=L/\ell$. Indeed, a generic eigenstate and the corresponding eigenvalue of the unperturbed Hamiltonian (\ref{E:H0}) have the form
\begin{equation}
\label{E:ges0}
|{\bf k},{\bf N}\rangle\rangle =\bigotimes_{c=1}^C |k_c, N_c\rangle_c, \quad E({\bf k},{\bf N})= \sum_c E_{k_c}^{N_c}
\end{equation}
 where ${\bf k} = (k_1,k_2,\ldots,k_C)$, ${\bf N} = (N_1,N_2,\ldots,N_C)$,  $ |k, N\rangle$ denotes the $k$th eigenstate of the cell Hamiltonian (\ref{E:cBHH}) for a population of $N$ bosons, and $E_k^{N}$ is the relevant eigenvalue. In the following we assume that these states are sorted according to their energy, so that the label $k=1$ refers to the ground state.
Thus, in particular, the ground state of $H_0$ relevant to the critical fractional filling $f=L/\ell$ (i.e. to the total population $N=CL$)  and the relevant energy have the form
\begin{equation}
\label{E:gs0m}
|\Psi_{C L}\rangle\rangle = \bigotimes_{c=1}^C |1, L\rangle_c, \quad {\cal E}^{(0)}_{C L} = C E_1^L
\end{equation}
Taking into account the translational symmetry of cell lattice, the {\it defect} ground states \cite{A:Freericks2} of $H_0$, corresponding to {\it particle} and {\it hole} total populations $N = C L \pm 1$, have the form \cite{N:dpt}
\begin{equation}
\label{E:gs0d}
|\Psi_{C L\pm1}\rangle\rangle = \frac{1}{\sqrt{C}}\sum_{j=1}^C |1, L\pm1\rangle_j \bigotimes_{c \neq j} |1, L\rangle_c, \quad {\cal E}^{(0)}_{C L\pm1} = (C-1) E_1^L+  E_1^{L\pm1}
\end{equation}
According to standard perturbative theory, the subsequent terms in the expansion of the ground state energies relevant to the total population ${\cal N}$
\begin{equation}
\label{E:gsE}
{\cal E}_{\cal N} = {\cal E}^{(0)}_{\cal N} + t_{\ell} {\cal E}^{(1)}_{\cal N}  + t_\ell^2 {\cal E}^{(2)}_{\cal N} 
\end{equation}
are
\begin{equation}
\label{E:E1}
{\cal E}^{(1)}_{\cal N} = \langle\langle \Psi_{\cal N}| V | \Psi_{\cal N} \rangle\rangle
\end{equation}
and
\begin{equation}
\label{E:E2}
{\cal E}^{(2)}_{\cal N} = {\sum_{{\bf k},{\bf N}}}'\frac{\langle\langle \Psi_{\cal N} | V | {\bf k},{\bf N}\rangle\rangle \langle\langle {\bf k},{\bf N}| V | \Psi_{\cal N}\rangle\rangle}{{\cal E}^{(0)}_{\cal N}-E({\bf k},{\bf N})} 
\end{equation}
where the prime signals that states with the same energy as the ground state must be excluded (recall that in the ``defect'' cases, ${\cal N} = C N \pm 1$, the ground state of $H_0$ is $C$-fold degenerate \cite{N:dpt}).

As it is discussed in some detail in the Appendix, the perturbative corrections  ${\cal E}_N^{(q)}$ are given in terms of matrix elements (on the unperturbed eigenbasis) of the single boson operators at the cell boundaries:
\begin{equation}
\label{E:MatEl}
\langle j,N  |a_k| h,N-1 \rangle,\qquad k \in \{1,\ell\}, 
\qquad N \in \{L-1,L, L+1,L+2 \} 
\end{equation}
Generally, such matrix elements must be evaluated numerically. However,  the necessary computational effort is easily affordable even for cells comprising several sites, and not only at the lowest critical fillings. Indeed the second order cell SCPE result for the domain boundary relevant to the critical filling $L/\ell$ of a $\ell$-periodic superlattice requires the complete diagonalization of a $\ell$-site cell BH Hamiltonian (\ref{E:cBHH}) for the five total populations ranging from $N=L-2$ to $N=L+2$. For instance, the most demanding operation to be faced when investigating the critical filling $11/5$ for a period 5 superlattice  is the complete diagonalization of a matrix whose size is $d(13,5)=2380$, which requires a negligible amount of time on a reasonably up-to-date personal computer. 

Note that, in order to obtain more refined results for the same system, one can resort to more complex numerical methods, such as quantum Monte Carlo simulations. However, the necessarily finite realization of the superlattice addressed by the simulation must comprise a sufficiently large number of cells to provide a good approximation of the thermodynamic limit. Since one expects that such number is of the order of the number of sites in a reasonably good finite-size realization of the one-dimensional homogeneous lattice, it is clear that even  superlattices with a small periodicity are computationally quite demanding.

The explicit form of Eqs.~(\ref{E:E1}) and (\ref{E:E2}) in terms of matrix elements in Eq.~(\ref{E:MatEl}) and the details of their derivation is given in the Appendix. Comparison with finite size exact results confirmed the correctness of our perturbative expansions.

\section{Results}
\label{S:results}
In this section the cell SCPE technique is adopted in the study of the insulator domain borders for several one dimensional superlattices. These borders are standardly obtained as $\mu_\pm = \pm [{\cal E}_{C L\pm 1} - {\cal E}_{C L}]$ and plotted in the $\mu/U$-$t/U$ plane, where we recall that $t$ is a scaling factor for the superlattice hopping amplitudes (i.e. $t_k = t\cdot \tau_k$). For a given value of $t$ we obtain the ground state energies ${\cal E}_{C L+ \sigma}$, with $\sigma=0,\pm 1$ from the second-order SCPE in Eq.~(\ref{E:gsE}). Notice that the perturbative order refers to the  parameter $t_\ell = t \tau_\ell$, and that the unperturbed Hamiltonian, Eq.~(\ref{E:cBHH}) depends on the overall hopping factor $t$ as well. This means that the perturbative estimates of the ground state energies and the insulator domain boundaries thereby are not second order functions of $t$.

\begin{figure}
\begin{center}
\includegraphics[width=8.5cm]{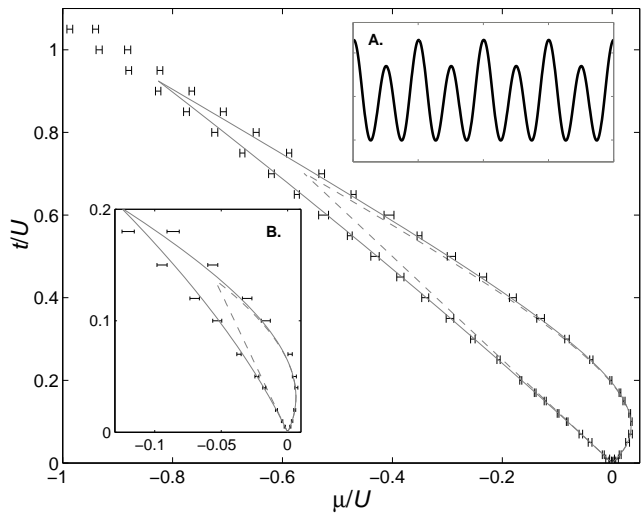}
\caption{\label{F:QMC} Half-filling insulator domain boundaries for a $\ell=2$ superlattice BH model with equal local potentials, $v_1=v_2$ (the corresponding optic potential is sketched in inset A). Main plot: $t_2 = 0.2 t_1 = 0.2 t$. Inset $t_2 = 0.6 t_1 = 0.6 t$. Solid lines: second order cell SCPE; Dashed lines: first order (analytic) cell SCPE \cite{A:loophole}; Errorbars: quantum Monte Carlo simulations on a superlattice comprising $C=50$ cells. }
\end{center}
\end{figure}

The simplest one-dimensional superlattice is the {\it dimerized chain}, corresponding to $\ell =2$ \cite{A:loophole,A:Bernstein74,A:Yamamoto}. As we mention, when $t_1=t_2$ the half-integer-filling insulating domains exhibit an unusual loophole shape. A qualitative description of these domains can be obtained resorting to a multiple-site mean-field approximation \cite{A:LoopLP}, whereas second order cell SCPE provides quite satisfactory quantitative results. This can be appreciated in Fig.~\ref{F:QMC}, where the domain borders of the half-filling insulating domain as obtained from cell SCPE and quantum Monte Carlo simulations are compared. One might expect  the perturbative technique to be valid in the limit $t_2 \ll t_1$. Note however that inset B of Fig.~\ref{F:QMC} shows that it  gives satisfactory results even for relatively large $t_2/t_1$.

\begin{figure}
\begin{center}
\includegraphics[width=8.5cm]{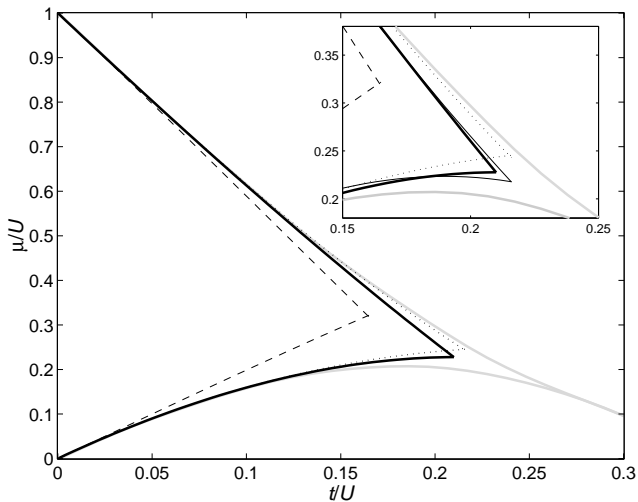}
\caption{\label{F:1D} First Mott lobe for the homogeneous 1D lattice according to: first-order, $\ell=2$ cell SCPE ( black dashed thin lines);  second-order, $\ell=2$  cell SCPE (black solid thick lines), standard third-order SCPE \cite{A:Freericks2} (black dotted thin lines);  density matrix renormalization group simulations \cite{A:Kuehner} (best known result, gray solid thick lines). The black solid thin line in the inset is the second order $\ell=4$ cell SCPE result.
  Note that it exhibits the reentrant character typical of the best result, unlike the standard third-order SCPE curves. }
\end{center}
\end{figure}

Actually, cell SCPE can be used to address the homogeneous lattice, setting $t_2=t_1 = t$. In this situation, the method correctly predicts the absence of fractional-filling loophole insulator domains, whereas the results for the integer filling domains are comparable to those provided by third-order standard SCPE \cite{A:Freericks2}, as it is shown in Fig.~\ref{F:1D}. Of course, the cell SCPE description of the homogeneous lattice can be performed considering larger ``unit cells''. In general, the thermodynamic convergence of the result does not seem to be very fast with increasing cell size. However, we notice that cells as small as four sites provide results exhibiting the expected reentrant character of the insulating domain (see inset of Fig.~\ref{F:1D}). 
Furthermore, the comparison between the $\ell=2$ first and second order cell SCPE results for the homogeneous lattice suggests that an improvement is more likely produced by an increase in the perturbative order rather than in the cell size.  In this respect, one might observe that cell SCPE are not really necessary in the homogeneous case. Still, as we discuss in the previous section, 
they are called for when dealing with non trivial periodic lattices.

As a further illustration of the cell SCPE technique, we address two superlattices with the same periodicity as the one realized experimentally by the authors of Ref.\cite{A:Peil}, i.e. $\ell =3$. We provide their zero-temperature phase diagrams considering insulating domains up to critical filling $7/3$. In this respect we mention that, in this particular case, cell SCPE would provide results for much larger fillings at a negligible computational cost. Conversely, the same results would be very demanding for more refined numerical methods, such as QMC  or density matrix renormalization group simulations.

\begin{figure}
\begin{center}
\includegraphics[width=8.5cm]{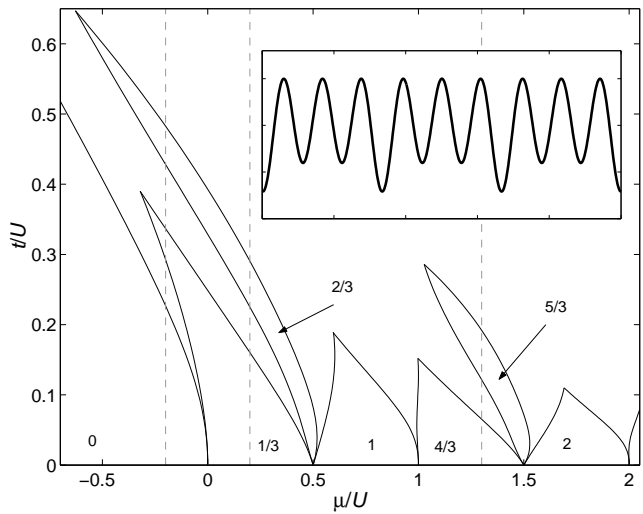}
\caption{\label{F:l3a} Zero-temperature phase diagram for a $\ell=3$ superlattice with  $t_1=t_2=t_3=t$, $v_1 = v_3 = 0.5 U$ and $v_2 = 0$ (the corresponding optic potential is sketched in the inset). The critical fillings characterizing each insulator domain are reported. The vertical dashed lines signal some values of the chemical potential where the reentrant behaviour of the insulator domains is evident (see the discussion in the text).}
\end{center}
\end{figure}

\begin{figure}
\begin{center}
\includegraphics[width=8.5cm]{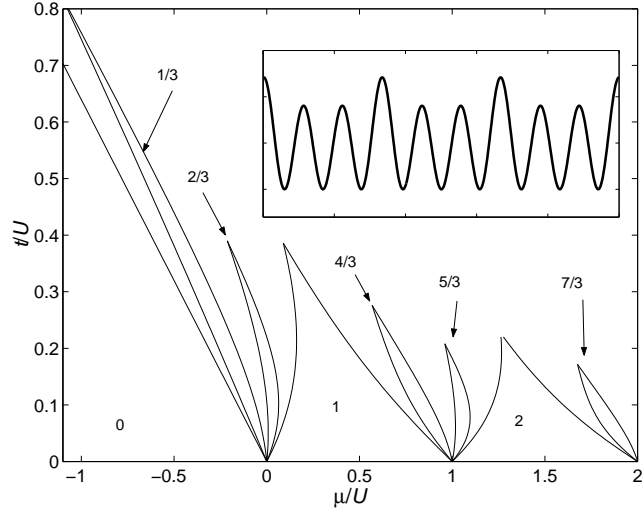}
\caption{\label{F:l3b} Zero-temperature phase diagram for a $\ell=3$ superlattice with  $t_1=t_2=t$, $t_3=0.3 t$ and $v_1 = v_2 = v_3 = 0$ (the corresponding optic potential is sketched in the inset). The critical fillings characterizing each insulator domain are reported. Also in the present case some of the insulating domains feature a reentrant shape. }
\end{center}
\end{figure}

The superlattice considered in Fig.~\ref{F:l3a} (sketched in the inset) is characterized by even hopping amplitudes ($t_1=t_2=t_3=t$) and local potentials $v_1 = v_3 = 0.5 U$ and $v_2 = 0$. As expected \cite{A:loophole}, the critical fillings of the form $f=(2+3 k)/3$ correspond to loophole insulator domains, while the remaining fillings yield usual lobe shaped domains. Figure~\ref{F:l3b} refers to a $\ell=3$ superlattice with even local potentials ($v_k=0$) hopping amplitudes $t_1 = t_2 = t$ and $t_3 = 0.3 t$. Again, cell SCPE provides the expected shape for the insulating domains, only the ones relevant to integer filling exhibiting the usual lobe shape \cite{A:loophole}.

The first striking feature emerging from the phase diagrams in Figs.~\ref{F:l3a} and \ref{F:l3b} is the size of the loophole insulator domains as compared to the usual Mott lobes. Note indeed that the mean-field results for the same systems predicts much smaller loophole domains  \cite{A:loophole,A:LoopLP}. Since SCPE provide a better quantitative estimate than mean-field, the quantum phase transition corresponding to loophole domains is expected to occur at not exceedingly small values of $t/U$.
Note furthermore that the reentrant character put into evidence in Ref.~\cite{A:Kuehner} for the first Mott lobe of the homogeneous one dimensional lattice is pervasive in the phase diagrams displayed in Figs.~\ref{F:l3a} and \ref{F:l3b}.
That is to say, values of the chemical potential $\mu$ exist where an increase in the hopping amplitude (i.e. in the kinetic energy) turns the system from superfluid to insulating. For instance in Fig.~\ref{F:l3a} this happens at $\mu = 1.3$, $\mu = 0.2$ and even twice at $\mu = -0.2$. This can be explained taking into account that an increase in $t/U$ at fixed $\mu/U$ can be achieved only by increasing the total boson population.
A maybe more experimentally relevant reentrant feature is that inherent in the loophole shape of some insulating domains. Note indeed that the width of an insulating domain, i.e. the chemical potential gap, is related to the excitation gap of the system at critical filling. An estimate of the latter, possibly achieved as described in Ref.~\cite{CM:Konabe}, would hence allow to discriminate loophole and lobe domains. Indeed the excitation gap is expected to decrease monotonically for the latter, while it features a maximum for the former.

\section{Discussion}
\label{S:concl}

In this paper we introduce the cell SCPE technique for the study of one-dimensional superlattice BH models. This technique provides a quantitative estimate of the zero-temperature phase diagram of the model at a low computational cost. Indeed we recall that  the data presented in Figs.~\ref{F:QMC}--\ref{F:l3b} are obtained in a few minutes of CPU time on a standard personal computer. Furthermore, cell SCPE provide a description of the insulator domain boundaries also for those critical fillings where standard perturbative approaches are ruled out, i.e. those corresponding to loophole domains \cite{A:loophole}. 

In Section~\ref{S:results}, The capabilities of our approach are illustrated by the zero-temperature phase diagrams of some experimentally realistic one-dimensional superlattices \cite{A:Peil}. We once again emphasize the fact that the loophole domains are entered at relatively large values of the parameter $t/U$ driving the quantum phase transition, which suggests that the observation of the latter does not require extreme experimental conditions.

With a reasonable effort, the technique --- described in some detail in the Appendix --- can be adapted to more complex  superlattices, such as the trimerized kagom\'e lattice whose experimental realization in terms of optic potentials is proposed in Ref.~\cite{A:Santos}. We emphasize that on such superlattices the computational demand of more refined numeric approaches, such as density matrix renormalization group and Quantum Monte Carlo simulations, can be overwhelming.
Conversely, the computational cost of second-order cell SCPE does not increase significantly for structures such as the kagom\'e lattice, and the only complication arises from the more complex topology of the cell lattice. This simply means that some more terms like those described in the Appendix have to be evaluated.
The study of the superfluid-insulator quantum phase transition for the BH model on more complex  superlattices, where the exact mapping with free spinless fermions does not apply, is the subject of a forthcoming paper \cite{N:forthcoming}.
\acknowledgments
The authors wish to thank V. Penna for fruitful discussion and H. Monien for kindly providing the DMRG data in Fig. 2.

% By means of cell SCPE, we evaluate the phase diagram of some relevant one-dimensional superlattices. We point out that the loophole domains appear for a large value of the parameter $t/U$, this suggests that they may be observed in experiments without requiring extreme experimental conditions. We remark that cell SCPE can be also applied to more complex superlattices such as the trimerized Kagom\'e which has been proposed in \cite{A:Santos} for an experimental realization. We note that on  larger dimensions the use of more refined techniques such as quantum Montecarlo and density matrix renormalization group is, in general, very demanding from a computational point of view. On such superlattices, the matrix elements in expressions (\ref{E:E1},\ref{E:E2}) have to be evaluate taking into account the topology of the new structure, and this may complicate to some extent the calculations, described in Appendix A. The study of the phase diagram for BH model on complex superlattices  will be the subject of a following paper.

\appendix*
\section{Details of the cell SCPE approach}
\label{S:details}
In this section we give a sketch of the derivation of the explicit form of the perturbative corrections in Eqs.~(\ref{E:E1}) and (\ref{E:E2}) in terms of matrix elements of the boson lattice operators at the cell boundaries, Eq.~(\ref{E:MatEl}).

Since the perturbative term $V$, Eq.~(\ref{E:perT}), is the sum of terms connecting neighbouring cells, the brackets in  Eqs.~(\ref{E:E1}) and (\ref{E:E2}) are the sum of terms like the following
\begin{eqnarray}
\label{E:prod}
\langle \langle {\bf k}',{\bf N}'| a^\dag_{\ell,j} a_{1,j+1}  |{\bf k},{\bf N}\rangle\rangle &=& 
_j\langle k'_j, N'_j|  a^\dag_{\ell,j}|k_j, N_j \rangle_j \cdot ~_{j+1}\langle k'_{j+1}, N'_{j+1}|  a_{1,j+i}|k_{j+1}, N_{j+1} \rangle_{j+1}\nonumber \\
&\cdot&  \prod_{c\neq j,j+1} ~_c\langle k'_c, N'_c| \mathbb{I}|k_c, N_c \rangle_c 
\end{eqnarray}
and its complex conjugate, where we recall that $|{\bf k},{\bf N}\rangle\rangle$, introduced in Eq.~(\ref{E:ges0}), denotes a generic eigenstate of the unperturbed Hamiltonian (\ref{E:cBHH}). Putting all of the above matrix elements in a stack, the previous product can be represented as
\begin{equation}
\label{E:stackG}
\begin{array}{c|c|c}
\vdots & \vdots & \vdots \\
\hline
k'_{j-1}, N'_{j-1} & {\mathbb I} & k_{j-1}, N_{j-1} \\
\hline
k'_j, N'_j & a_\ell^\dag & k_{j}, N_{j} \\
\hline
k'_{j+1}, N'_{j+1} & a_1 & k_{j+1}, N_{j+1} \\
\hline
k'_{j+2}, N'_{j+2} & {\mathbb I} & k_{j+2}, N_{j+2} \\
\hline
\vdots & \vdots & \vdots 
\end{array}
\end{equation}
where the central column pertains to the operator term, while the lateral columns represent the states involved in the product. Each level of the stack refers to a given cell of the superlattice, and contains the corresponding term of the product in Eq.~(\ref{E:prod}). 
Of course, the product in the rows containing an identity operator vanishes unless $k'=k$, $N'=N$, in which case it equals one. Likewise the rows containing the boson operators vanish unless $N' = N\pm 1$,  but in general there is no constraint on the energy labels $k$ and $k'$. Note that we dropped the cell labels in the boson operators appearing in the stack.
Henceforth we drop the cell labels unless strictly necessary.

We are now ready to evaluate the first order terms of the energy expansions. When the total population in the superlattice is  ${\cal N} = C L$, where $C$ is the total number of cells and $L$ is an integer, the pictorial representation of the terms in Eq. (\ref{E:E1}) is
\begin{equation}
\begin{array}{c|c|c}
\vdots & \vdots & \vdots \\
\hline
1, L & {\mathbb I} & 1,L \\
\hline
1, L & a_\ell^\dag & 1,L \\
\hline
1,L & a_1 & 1,L \\
\hline
1,L & {\mathbb I} & 1,L \\
\hline
\vdots & \vdots & \vdots 
\end{array}
\end{equation}
According to the previous discussion, all of these term vanish. Hence ${\cal E}_{C L}^{(1)}=0$.

When ${\cal N} = C L+1$ it is clear that the only non-vanishing terms are of the form
\begin{equation}
\begin{array}{c|c|c}
\vdots & \vdots & \vdots \\
\hline
1, L & {\mathbb I} & 1,L \\
\hline
1, L+1 & a_\ell^\dag & 1,L \\
\hline
1,L & a_1 & 1,L+1 \\
\hline
1,L & {\mathbb I} & 1,L \\
\hline
\vdots & \vdots & \vdots 
\end{array}
\label{stack1}
\end{equation}
and its complex conjugate.
More precisely, two of such terms are contributed by  each of the $C$ cells. Taking into account the normalization factor of the symmetrized ground state $|\Psi_{C L+1}\rangle\rangle$, Eq.~(\ref{E:gs0d}), one gets
\begin{eqnarray}
{\cal E}_{C L+1}^{(1)} & = & \langle 1, L+1| a_\ell^\dag |1,L \rangle  \langle 1, L| a_1 |1,L+1 \rangle +
\langle 1, L| a_\ell |1,L+1 \rangle  \langle 1, L+1| a^\dag_1 |1,L \rangle \nonumber \\
& = & 2 \cdot \langle 1, L+1| a_\ell^\dag |1,L \rangle  \langle 1, L| a_1 |1,L+1 \rangle 
\end{eqnarray} 
where in the second equation we take advantage of the fact that the operators and the states are real.
Likewise
\begin{equation}
{\cal E}_{C L-1}^{(1)}= 2 \langle 1, L| a_\ell^\dag |1,L-1 \rangle  \langle 1, L-1| a_1 |1,L \rangle 
\end{equation} 

Let us now address the second order term in the cell SCPE, Eq.~(\ref{E:E2}). Each term in the corresponding sum contains a product two matrix elements of the the form (\ref{E:prod}). However, most of these product vanish.  
In order to single out the non-vanishing products we adopt a trivial generalization of the stack representation introduced above. The stack representing the product under concern is simply obtained placing two stacks of the form (\ref{E:stackG}) side by side. According to Eq.~(\ref{E:E2}), the innermost columns of such six-column stack refer to the same unperturbed state. The outermost column refer once again to the ground state, or more precisely to one of the possibly degenerate unperturbed states appearing in its expression, Eqs.~(\ref{E:gs0m}) and ~(\ref{E:gs0d}).

When ${\cal N} = C L$, the unperturbed ground state is a single state of the Fock space, Eq.~(\ref{E:gs0m}). The corresponding stack representation for the second order correction to the ground state energy is
\begin{equation}
\begin{array}{c|c|c|c|c|c}
\vdots & \vdots & \vdots & \vdots & \vdots & \vdots \\
\hline
1, L & {\mathbb I} & 1,L & 1, L & {\mathbb I} & 1,L \\
\hline
1, L & a_\ell  & j,L+1 & j, L+1 & a_\ell^\dag & 1,L \\
\hline
1, L & a_1^\dag  & h,L-1 & h,L-1 & a_1 & 1,L \\
\hline
1, L & {\mathbb I} & 1,L & 1,L & {\mathbb I} & 1,L \\
\hline
\vdots & \vdots & \vdots & \vdots & \vdots & \vdots 
\end{array}
\label{CL}
\end{equation}
By inserting the matrix elements corresponding to stack (\ref{CL}) in Eq.~(\ref{E:E2}) one gets
\begin{equation}
\label{E:CL2}
{\cal E}_{C L}^{(2)}=C \sum_{j=1}^{D_{L+1}} \sum_{h=1}^{D_{L-1}} \frac{\left| \langle j,L+1|a_\ell^\dag | 1,L\rangle \langle h,L-1|a_1 | 1,L\rangle \right|^2+ \left| \langle j,L+1|a_1^\dag | 1,L\rangle \langle h,L-1|a_\ell | 1,L\rangle \right|^2
}{2 E_1^L - E_j^{L+1}- E_h^{L-1}}
\end{equation}
where the first term in the numerator corresponds to the process illustrated in stack (\ref{CL}), the second to its complex conjugate and $D_L = d(L,\ell)$ is the size of the Fock space for $L$ bosons on $\ell$ sites.
 The overall factor  $C$ takes into account the fact that the operator $V$ is obtained as the sum of $C$ terms.

Let us now consider the case where the total number of bosons is  ${\cal N} = C L$. The unperturbed ground  state is now given by the superposition of Fock states in Eq.~(\ref{E:gs0d}). This gives rise to five different contributions whose stack representations are discussed below.

First of all we notice that there is a stack giving a contribution of the form  (\ref{E:CL2}), yet with a different overall factor. This corresponds to
\begin{equation}
% \begin{center}
\begin{array}{c|c|c|c|c|c}
\vdots & \vdots & \vdots & \vdots & \vdots & \vdots \\
\hline
1, L+1 & {\mathbb I} & 1,L+1 & 1, L+1 & {\mathbb I} & 1,L+1 \\
\hline
1, L & {\mathbb I} & 1,L & 1, L & {\mathbb I} & 1,L \\
\hline
1, L & a_\ell  & j,L+1 & j, L+1 & a_\ell^\dag & 1,L \\
\hline
1, L & a_1^\dag  & h,L-1 & h,L-1 & a_1 & 1,L \\
\hline
1, L & {\mathbb I} & 1,L & 1,L & {\mathbb I} & 1,L \\
\hline
\vdots & \vdots & \vdots & \vdots & \vdots & \vdots 
\end{array}
% \end{center}
\label{CL+1a}
\end{equation}
producing a contribution of the form
\begin{equation}
{\cal E}_{C L+1,a}^{(2)}=(C-2) \sum_{j=1}^{D_{L+1}} \sum_{h=1}^{D_{L-1}} \frac{\left| \langle j,L+1|a_\ell^\dag | 1,L\rangle \langle h,L-1|a_1 | 1,L\rangle \right|^2+ \left| \langle j,L+1|a_1^\dag | 1,L\rangle \langle h,L-1|a_\ell | 1,L\rangle \right|^2
}{2 E_1^L - E_j^{L+1}- E_h^{L-1} }
\end{equation}
where the factor $C-2$ takes into account the number of terms in $V$ giving rise to the processes described by the stack (\ref{CL+1a}).

Another contribution corresponds to the stack
\begin{equation}
% \begin{center}
\begin{array}{c|c|c|c|c|c}
\vdots & \vdots & \vdots & \vdots & \vdots & \vdots \\
\hline
1, L & {\mathbb I} & 1,L & 1, L & {\mathbb I} & 1,L \\
\hline
1, L+1 & {\mathbb I}  & j,L+1 & j, L+1 & a_\ell^\dag & 1,L \\
\hline
1, L & a_\ell^\dag  & h,L-1 & h,L-1 & a_1 & 1,L \\
\hline
1, L & a_1 & j',L+1 & j',L+1 & {\mathbb I} & 1,L+1 \\
\hline
1, L & {\mathbb I} & 1,L & 1, L & {\mathbb I} & 1,L \\
\hline
\vdots & \vdots & \vdots & \vdots & \vdots & \vdots 
\end{array}
% \end{center}
\label{CL+1b}
\end{equation}
that vanishes unless  $j=j'=1$. This produces  a contribution of the form
\begin{eqnarray}
{\cal E}_{C L+1,b}^{(2)} & = &  \sum_{h=1}^{D_{L-1}} 
\frac{ \langle 1,L+1|a_\ell^\dag | 1,L\rangle \langle 1,L|a_\ell^\dag | h,L-1\rangle 
\langle h,L-1|a_1 | 1,L\rangle \langle 1,L|a_1 | 1,L+1\rangle} 
{2 E_1^L - E_1^{L+1}- E_h^{L-1} } \nonumber\\
& + & \sum_{h=1}^{D_{L-1}} 
\frac{ \langle 1,L+1|a_1^\dag | 1,L\rangle \langle 1,L|a_1^\dag | h,L-1\rangle 
\langle h,L-1|a_\ell | 1,L\rangle \langle 1,L|a_\ell | 1,L+1\rangle} 
{2 E_1^L - E_1^{L+1}- E_h^{L-1} } \nonumber\\
& = &  2 \cdot \sum_{h=1}^{D_{L-1}} 
\frac{ \langle 1,L+1|a_\ell^\dag | 1,L\rangle \langle 1,L|a_\ell^\dag | h,L-1\rangle 
\langle h,L-1|a_1 | 1,L\rangle \langle 1,L|a_1 | 1,L+1\rangle} 
{2 E_1^L - E_1^{L+1}- E_h^{L-1} }
\end{eqnarray}
where the second term is obtained by  complex conjugation. 

The third stack to be considered is
\begin{equation}
% \begin{center}
\begin{array}{c|c|c|c|c|c}
\vdots & \vdots & \vdots & \vdots & \vdots & \vdots \\
\hline
1, L & {\mathbb I} & 1,L & 1, L & {\mathbb I} & 1,L \\
\hline
1, L & a_\ell  & j,L+1 & j, L+1 & a_\ell^\dag & 1,L \\
\hline
1, L+1 & a_1^\dag & h,L & h,L & a_1 & 1,L+1 \\
\hline
1, L & {\mathbb I} & 1,L & 1, L & {\mathbb I} & 1,L \\
\hline
\vdots & \vdots & \vdots & \vdots & \vdots & \vdots 
\end{array}
% \end{center}
\label{CL+1c}
\end{equation}
where  $j= h=1$ must be discarded since in this case the innermost vectors  belong to the lower eigenspace of $H_0$. The process described by the stack (\ref{CL+1c}) and its complex conjugate produce the contribution
\begin{equation}
{\cal E}_{C L+1,c}^{(2)}={\sum_{j=1}^{D_{L+1}}} {\sum_{h=1}^{D_{L}}} \sigma_{j,h} \frac{\left| \langle j,L+1|a_\ell^\dag | 1,L\rangle \langle h,L|a_1 | 1,L+1\rangle \right|^2+ \left| \langle j,L+1|a_1^\dag | 1,L\rangle \langle h,L|a_\ell | 1,L\rangle \right|^2
}{E_1^L+E_1^{L+1} - E_j^{L+1}- E_h^{L} }
\end{equation}
where $\sigma_{j h}$ equals zero if $j=h=1$ and one otherwise.

The next stack we consider is
\begin{equation}
% \begin{center}
\begin{array}{c|c|c|c|c|c}
\vdots & \vdots & \vdots & \vdots & \vdots & \vdots \\
\hline
1, L & {\mathbb I} & 1,L & 1, L & {\mathbb I} & 1,L \\
\hline
1, L+1 & a_\ell^{\dag}  & h',L & h', L & {\mathbb I} & 1,L \\
\hline
1, L & a_1 & j,L+1 & j, L+1 & a_\ell^{\dag} & 1,L \\
\hline
1, L & {\mathbb I} & h,L & h,L & a_1 & 1,L+1 \\
\hline
1, L & {\mathbb I} & 1,L & 1, L & {\mathbb I} & 1,L \\
\hline
\vdots & \vdots & \vdots & \vdots & \vdots & \vdots 
\end{array}
% \end{center}
\label{CL+1d}
\end{equation}
which produces a non-vanishing contribution only if $h=h'=1$. Furthermore  $j=1$ must be discarded,  since in this case the innermost vectors  belong to the lower eigenspace of $H_0$. Therefore, one gets
\begin{eqnarray}
{\cal E}_{C L+1,d}^{(2)} & = &  \sum_{j=2}^{D_{L+1}} 
\frac{ \langle 1,L+1|a_\ell^\dag | 1,L\rangle \langle 1,L|a_1 | j,L+1\rangle 
\langle j,L+1|a_\ell^\dag | 1,L\rangle \langle 1,L|a_1 | 1,L+1\rangle} 
{E_1^{L+1}- E_j^{L+1} } \nonumber\\
& + & \sum_{j=2}^{D_{L+1}} 
\frac{ \langle 1,L+1|a_1^\dag | 1,L\rangle \langle 1,L|a_\ell | j,L+1\rangle 
\langle j,L+1|a_1^\dag | 1,L\rangle \langle 1,L|a_\ell | 1,L+1\rangle} 
{E_1^{L+1}- E_j^{L+1} } \nonumber\\
&=& 2\cdot \sum_{j=2}^{D_{L+1}} 
\frac{ \langle 1,L+1|a_\ell^\dag | 1,L\rangle \langle 1,L|a_1 | j,L+1\rangle 
\langle j,L+1|a_\ell^\dag | 1,L\rangle \langle 1,L|a_1 | 1,L+1\rangle} 
{E_1^{L+1}- E_j^{L+1} }
\end{eqnarray}

Finally, the stack
\begin{equation}
% \begin{center}
\begin{array}{c|c|c|c|c|c}
\vdots & \vdots & \vdots & \vdots & \vdots & \vdots \\
\hline
1, L & {\mathbb I} & 1,L & 1, L & {\mathbb I} & 1,L \\
\hline
1, L & a_\ell  & j,L+2 & j, L+2 & a_\ell^\dag & 1,L+1 \\
\hline
1, L+1 & a_1^\dag & h,L-1 & h,L-1 & a_1 & 1,L \\
\hline
1, L & {\mathbb I} & 1,L & 1, L & {\mathbb I} & 1,L \\
\hline
\vdots & \vdots & \vdots & \vdots & \vdots & \vdots 
\end{array}
% \end{center}
\label{CL+1e}
\end{equation}
produces the contribution
\begin{equation}
{\cal E}_{C L+1,e}^{(2)}={\sum_{j=1}^{D_{L+2}}} {\sum_{h=1}^{D_{L-1}}} \frac{\left| \langle j,L+2|a_\ell^\dag | 1,L+1\rangle \langle h,L-1|a_1 | 1,L\rangle \right|^2+ \left| \langle j,L+2|a_1^\dag | 1,L\rangle \langle h,L-1|a_\ell | 1,L\rangle \right|^2
}{E_1^L+E_1^{L+1} - E_j^{L+2}- E_h^{L-1} }
\end{equation}

Putting all together, the second-order correction to the ground-state energy, Eq.~(\ref{E:E2}), is
\begin{equation}
{\cal E}_{C L+1}^{(2)}={\cal E}_{C L+1,a}^{(2)}+{\cal E}_{C L+1,b}^{(2)}+{\cal E}_{C L+1,c}^{(2)}+
{\cal E}_{C L+1,d}^{(2)}+{\cal E}_{C L+1,e}^{(2)}
\end{equation}

%\bibliographystyle{../../BIBTEX/etal10}
%\bibliography{../../BIBTEX/biblio,./ricopN}

\begin{thebibliography}{10}
\bibitem{N:dpt}
This can be made rigorous resorting to degenerate perturbative theory. Note
  indeed that all of the $C$ states summed up in Eq.~(\ref{E:gs0d}) belong to
  the same (degenerate) eigenspace of $H_0$.
\end{thebibliography}

\end{document}